\documentclass[sn-mathphys]{sn-jnl}

\jyear{2022}%
\usepackage{amsmath}

\begin{document}

\title[Dynamical information flow within the MI system]{Dynamical information flow within the magnetosphere-ionosphere system during magnetic storms}

\author*[1,2]{\fnm{Mirko} \sur{Stumpo}}\email{mirko.stumpo@inaf.it}

\author[1]{\fnm{Simone} \sur{Benella}}\email{simone.benella@inaf.it}

\author[1]{\fnm{Giuseppe}\sur{Consolini}}\email{giuseppe.consolini@inaf.it}

\author[1]{\fnm{Tommaso} \sur{Alberti}}\email{tommaso.alberti@inaf.it}

\affil*[1]{\orgdiv{INAF - Istituto di Astrofisica e Planetologia Spaziali}, \city{Roma}, \country{Italy}}

\affil[2]{\orgdiv{Dipartimento di Fisica}, \orgname{Università degli Studi di Roma Tor Vergata}, \city{Roma}, \country{Italy}}


\abstract{The direct role of successive intense magnetospheric substorms in injecting/energizing particles into the storm-time ring current is still debated and controversial. Whereas in the recent past it has been observed the absence of a net information flow between magnetic storms and substorms, previous \textit{in-situ} satellite observations have evidenced that ionospheric-origin ions dominate the population of the ring current during the main phase of geomagnetic storms. As a matter of fact, the controversy arises mainly by the use of sophisticated data-driven techniques somewhat contradicting \textit{in-situ} measurements. In this framework, the main aim of this work is to attempt an adaption of the powerful information-theoretic approach, i.e., the transfer entropy, in a consistent way with physics modeling and observations and to explore the possible motivations behind the underlying contradictions that emerge when these techniques are used. Our idea is to characterize the dynamics of the information flow within the magnetosphere-ionosphere system using a database of geomagnetic storms instead of considering a long time series of geomagnetic indices. This allows us to consider local non-stationary features of the information flow and, most importantly, to follow the transition from quiet to disturbed periods and vice-versa.}


\maketitle

\section{Introduction}\label{sec1}

When the interplanetary magnetic field (IMF) is characterized by a nearly-southward orientation for a sufficiently long time, the near-Earth electromagnetic environment, i.e., the plasma circulation and the magnetospheric/ionospheric current systems, undergoes some dynamical changes to dissipate the excess of energy, momentum and mass coming from the surrounding interplanetary medium \citep{akasofu1961ring,akasofu1974auroral}. As a consequence, magnetic storms and substorms develop, being the macroscopic manifestation of these dynamical processes of dissipation\citep{gonzalez1994geomagnetic}. 

In recent years, understanding the physical mechanisms behind the development of these events is becoming more and more important since magnetic disturbances may be tremendously hazardous for telecommunications, satellites preservation and also responsible for exposing astronauts to abnormal radiation level \citep{Malandraki}. The fingerprint of a magnetic storm is the depression of the horizontal component of the magnetic field caused by the enhancement of the ring current, usually monitored by the Disturbance Storm Time ($D_{st}$) index (1 hour resolution by definition) or the SYM-H index, which is its 1-minute resolution equivalent. It is now recognized that the efficiency of energization is, in turn, primarily controlled by a long-lasting southward component of the IMF. However, whether the southward IMF is the unique driver of both particle energization and injection or if other magnetospheric-ionospheric internal processes are responsible for large scale injection into the storm-time ring current is still an open problem \citep{borovsky2021our}. 

Since the pioneering works by Akasofu and Chapman \citep{akasofu1961ring}, magnetospheric substorms (i.e., violent electrojet activity) have been identified as a possible source of particle injection, mainly because of their statistical association with the occurrence of the magnetic storm main phase \citep{kamide1998current}. In particular, based on investigating particle injection at geostationary orbit it was concluded that substorms drive particle acceleration \citep{akasofu1974auroral}. Soon after, it has been shown that the constituents of the storm-time ring current are more energetic than those energized directly by substorms \citep{williams1987ring}. On the other hand, predictions of the $D_{st}$ index, by using the auroral index AL alone, are well in agreement with observations \citep{burton1975empirical, kamide1971analysis, gonzalez1994geomagnetic}. Furthermore, AMPTE and CRRES missions \citep{krimigis1982active, wilken1992magnetospheric} also reported the direct observation of ionospheric-origin ions dominating the population of the ring current during the main phase of a storm \citep{hamilton1988ring, daglis1997role}, suggesting a direct causal link with substorms. Indeed the upward acceleration along the magnetic field lines of these ionospheric ions may be associated to successive occurrence of intense substorms. 

Despite such important observations, it remains extremely difficult to disentangle the role of substorms in the energization of the storm-time ring current directly from geomagnetic indices data.
First attempts have been made by using prediction filters and, more recently, by using sophisticated techniques based on information theory \citep{de2011information, stumpo2020measuring, manshour2021causality}. All these works share the common idea that causation is associated to the notion of predictability, instead of correlation. If the knowledge of the time series $Y$ reduces the error in predicting the time series $X$ it is said that an information flow (IF; or, equivalently, predictive causality) exists from $Y$ to $X$. This is the reason why predictions are associated to evidence of asymmetric couplings between the components of a physical system (in this case the magnetosphere-ionosphere system).

The main drawback in using data-driven techniques is conceptual. Can geomagnetic indices as $D_{st}$ (or SYM-H) and AL capture the processes of injection and energization \citep{mcpherron1997role}? Other issues arise when conditional statistics are used. Indeed, recently Runge et al. \citep{runge2018common} and Manshour et al. \citep{manshour2021causality} have shown, using conditional transfer entropy, that if one removes the influence of the southward IMF, the IF from AL/AE to SYM-H (and vice-versa) previously found by De Michelis et al. \citep{de2011information} and Stumpo et al. \citep{stumpo2020measuring} becomes negligible. These studies are carried out, respectively, by using 20 minutes time averaged SYM-H and 5 minutes resampled SYM-H, although McPherron \citep{mcpherron1997role} suggested that high-resolution indices are required for unveiling the role of substorms because otherwise fast dynamics may be lost in average. On the other hand, other authors suggested that a magnetic storm is not a trivial superposition of intense substorms, but that the outflow of ionospheric ions is controlled by an efficiency function $\alpha=\alpha(B_z, t)$ which depends on the southward IMF. In particular, according to Kamide et al. \citep{kamide1992substorm} and Gonzalez et al. \citep{gonzalez1994geomagnetic}, the energy balance equation for $D_{st}$ (justified in physics ground by the Dessler-Parker-Sckopke relation) can be written as
\begin{equation}
    \frac{d}{dt}D_{st}(t) = \alpha(B_z, t) \text{AL}(t) - \frac{D_{st}(t)}{T},
    \label{energy_balance}
\end{equation}
where $T$ is the relaxation time of $D_{st}$, i.e., the duration of the disturbance, which in general is not constant but depends on the specific concentrations of ions species in the ring current. Now, IF the southward $B_z$ controls the efficiency, i.e., the energy supply, when $B_z$ is removed from the model, the IF from AL to $D_{st}$ may disappear. Equation \eqref{energy_balance} suggests also that the IF from AL to $D_{st}$ is non-negligible only during southward $B_z$, also according to observations of particle injection discussed above. This fact implies that the IF is strongly non-stationary and, as such, it is different during storm and non-storm times. Thus, unraveling the IF by using long time series to estimate the average of transfer entropy may be misleading, since $\alpha(B_z, t)$ is reasonably different from zero only for short times. As a result, the use of long time series most likely tends to unbalance the contributions of quiet and disturbed geomagnetic periods, thus leading to bias the IF towards values mainly due to non-storm time intervals. This argument may be at the origin of the vanishing IF observed by Runge et al. \citep{runge2018common} and Manshour et al. \citep{manshour2021causality}.

In this framework, the aim of this paper is to shed some light in the debate about the influence of magnetospheric substorms on the storm-time ring current development. The main task is to suit the powerful information-theoretic approach to the study of this issue in a consistent way with physics modeling and in-situ observations. To achieve this we consider a dataset of storms and substorms extracted from SuperMAG database \citep{gjerloev2009global} and we perform the estimation of the transfer entropy between high-latitude, low-latitude geomagnetic indices and IMF $B_z$ component by using a sliding window technique. 
Since small time windows contain non-stationary data 
we consider a set of storm events, i.e. an ensemble of independent time series, that allows us to investigate the dynamics of the IF between different physical quantities during the storm events. In fact, the main advantage of this approach is that quiet and disturbed periods are not averaged in the analysis and thus the ensemble enables to quantify how the IF varies during the different phases of the storm. 

The paper is organized as follows: in Section \ref{sec:2} we will review the concept of IF and the definition as well as the derivation of the transfer entropy. Next we will discuss the motivation behind the use of this tool as well as advantages and disadvantages of the ensemble approach for geomagnetic studies. In Section \ref{sec:3} we present the dataset and we discuss the synchronization of signals involved in the ensemble. Finally in Section \ref{sec:4} we show the results, while in Section \ref{sec:5} we discuss them in terms of interpretation and contextualization in literature. 

\section{Methods}\label{sec:2}
Generally, the study of causation is dominated by the notion of predictability \citep{sugihara2012detecting}. In this context, if we have two time series $X$ and $Y$, we say that $Y$ drives the dynamics of $X$ if the information about the state of $X$ can be recovered from the past states of $Y$ and not vice-versa. 
In the framework of non-parametric statistics, the formalization of this concept can be retrieved in the notion of IF.

The step forward of the IF with respect to the popular concept of correlation is that the former removes any redundant or shared information between current $X$ and its own past. For example,
if we have two processes, $X$ and $Y$, such that $Y$ drives $X$ and not vice-versa, we need to consider that the process $Y_t$ incorporates intrinsically the information about $X_t$'s past, otherwise the cross correlation (or, equivalently, the delayed mutual information) in the direction from $X$ to $Y$ would not be zero even if the IF is absent. For further details see \citep{bossomaier2016transfer} and \citep{schreiber2000measuring}.

Therefore, if we want to compute the IF from $Y$ to $X$, the idea is to remove the redundancy introduced by the past history of $X$. To account this, the most simple formulation of transfer entropy is given in terms of conditional mutual information (CMI), i.e.
\begin{equation}\label{transfer_entropy_1}
    T_{Y\rightarrow X}^{(k,l)}(\tau) = I(X_t; \mathbf{Y}_{t-\tau}^{(l)} \vert \mathbf{X}_{t-1}^{(k)}),
\end{equation}
where in general $\mathbf{Y}_{t-\tau}^{(l)} = \left(Y_{t-\tau}, Y_{t-\tau-1}, ... , Y_{t-\tau-l} \right)$ and $\mathbf{X}_{t}^{(k)} = \left(X_t, X_{t-1}, ... , X_{t-k} \right)$ are the multivariate reconstruction of $Y$ and $X$ past histories, respectively, if we assume they are $l$-th and $k$-th order Markov processes. From a probabilistic point-of-view, the transfer entropy is essentially the distance in probability from the validity of the generalized Markov condition, i.e. 
\begin{equation}\label{generalMarkov}
    p(X_{t}\vert \mathbf{X}_{t-1}^{(k)};\mathbf{Y}_{t-\tau}^{(l)})=p(X_{t}\vert\mathbf{X}_{t-1}^{(k)}),
\end{equation}
which is fulfilled if and only if $X_t$ depends (conditionally) only on its own history. Using conditional Kullback Leibler Divergence (cKLD), we can compute the distance between l.h.s and r.h.s of eq. \eqref{generalMarkov} and obtain the explicit formula for the transfer entropy \citep{schreiber2000measuring}
\begin{equation}\label{transfer_entropy_2}
    T_{Y \rightarrow X}^{(k,l)}(\tau) = \sum_{X_{t},\mathbf{X}_{t-1}^{(k)},\mathbf{Y}_{t-\tau}^{(l)}} p(X_{t},\mathbf{X}_{t-1}^{(k)},\mathbf{Y}_{t-\tau}^{(l)})\log\frac{p(X_{t}\vert \mathbf{X}_{t-1}^{(k)},\mathbf{Y}_{t-\tau}^{(l)})}{p(X_{t}\vert \mathbf{X}_{t-1}^{(k)})}.
\end{equation}
To compute eq. \ref{transfer_entropy_2} the KSG estimator is used for its optimality in terms of systematic errors and biases due to finite sample effects \citep{kraskov2004synchronization, kraskov2004estimating, wibral2014transfer}.

Note that whereas in general for non-Markov process both $(k, l) \rightarrow (\infty, \infty)$, for purely Markov processes, the past history of e.g. $X_t$ is completely embedded into $X_{t-1}$ alone, so that the redundancy of $X$'s own past can be elimated by simply conditioning on $X_{t-1}$ and the multivariate vector $\mathbf{X}_{t-1}^{(k)}$ collapses to the univariate signal $\mathbf{X}_{t-1}^{(k)} \rightarrow X_{t-1}$. At this point it is also worth noticing that $\mathbf{X}_{t-1}^{(k)}$ and $\mathbf{Y}_{t-\tau}^{(l)}$ can be thought as an embedding reconstruction of the phase space according to Taken's theorem \citep{takens1981detecting}, although the equivalence with the real phase-space is not guaranteed for stochastic systems \citep{kantz2004nonlinear}. Furthermore, when the noise-level is high, or when the underlying dynamics is stochastic at coarse grained scales, the parameters of the reconstruction $k$ and $l$, cannot be recovered by using methods, such as False Nearest Neighbours technique \citep{kennel1992determining}, because they are suited for deterministic dynamics \citep{ragwitz2002markov,kantz2004nonlinear}. 

Our definitions in eqs. \eqref{transfer_entropy_1} and \eqref{transfer_entropy_2} incorporate directly the time lag $\tau$ between $X$ and $Y$. This is because the interaction may be delayed in time more than $\tau=1$ as in the original definition by Schreiber \citep{schreiber2000measuring}. In this case, only $Y$ is lagged forward in time, while the past of $X$ remains untouched. This choice, as demonstrated explicitly by Wibral et al. \citep{wibral2013measuring}, is the only one, among the most popular definitions of information transfer (e.g. \citep{pompe2011momentary}, \citep{paluvs2001synchronization}), allowing to restore Wiener's principle of causality and to recover the correct transfer delays \citep{bossomaier2016transfer}, \citep{wibral2013measuring}.

Practically, the interpretation of the transfer entropy is straightforwardly related to the difference in the uncertainties of $X$'s future before and after the knowledge of $Y$'s past, recovering again the notion of predictability in the statistical inference of IF. Following \citep{bossomaier2016transfer} this interpretation can be put in a formal way by decomposing the total uncertainty, i.e. the Shannon entropy, $H(X_t)$ as
\begin{equation}\label{information_balance}
    H(X_t) = I(\mathbf{X}_{t-1}^{(k)}; X_{t}) + T_{Y \rightarrow X}^{(k,l)}(\tau) + H(X_t \vert \mathbf{X}_{t-1}^{(k)}; \mathbf{Y}_{t-\tau}^{(l)}).
\end{equation}
Hence, we can extract three contributions in the uncertainty on the future state of $X$: the first accounts for the information contained in the past $\mathbf{X}_{t-1}^{(k)}$, the second term is the IF from the $Y$'s past, and the third term is the residual uncertainty after the knowledge of both $X$'s and $Y$'s histories. From eq. \eqref{information_balance} it is clear that the consideration of $\mathbf{X}_{t-1}^{(k)}$ instead of $X_{t-1}$ controls the balance between stored and transferred information. An inadequate reconstruction of the k-th Markov process may lead to confusion between stored and transferred information \citep{bossomaier2016transfer}.

In general, when the information-theoretic approach is applied to real data, we have single realizations of $X$ and $Y$ as time series processes. In this case, the PDFs in eq. \eqref{transfer_entropy_2} are estimated assuming stationarity and, naturally, also the IF in this case should be stationary. The consequence is that transient dynamics and local non-stationarity is completely neglected. This point is crucial when the aim is to study the role of intense substorms in the energization of storm-time ring current. 
For this reason, as mentioned in the Introduction, we need a time-resolved estimation of the transfer entropy over a sliding window as presented below. 

Let $W=\{\hat{t}-\delta, \hat{t}-\delta+1, ..., \hat{t}+\delta \}$ be the time window centered around $\hat{t}$. Then, we can restrict the time series $X_t$, $\mathbf{X}_{t-1}^{(k)}$ and $\mathbf{Y}_{t-\tau}^{(l)}$ to $W$ and compute eq. \eqref{transfer_entropy_2} to get the time windowed transfer entropy, i.e.
\begin{equation}
    \mathcal{T}_{Y\rightarrow X}^{(k, l)}(\hat{t}, \tau) = T_{Y_W \rightarrow X_W}^{(k,l)}(\tau).
\end{equation}
In principle, in the limit $N\to \infty$ and $\delta \to 0$, the average $\langle \mathcal{T}_{Y\rightarrow X}^{(k, l)}(\hat{t}, \tau) \rangle_{\hat{t}}$ converges to eq. \eqref{transfer_entropy_2}.
As can be seen from this definition, one of the main limitation of the method when we deal with empirical observations, is represented by the need for a sufficient statistics, which cannot be always guaranteed. Indeed, geomagnetic indices are sampled at a maximum cadence of 1 minute and the used time window must be large enough to include a sufficient number of data points in the statistics. On the other hand, this size cannot be too large either since we want to resolve transient dynamics, which could be suppressed by averaging on wide time windows.

In order to overcome this limitation, firstly we find a suitable trade-off for the size of the time window used in the analysis and then we introduce the estimation of the transfer entropy over an ensemble of independent realizations of magnetic storms in a way similar to the method proposed by Gómez-Herrero et al. \citep{gomez2015assessing}. This enables us to study how the IF varies during the evolution of the magnetic storm. We remark that the ensemble approach is somewhat different from computing transfer entropy between individual trials averaging the single transfer entropies \textit{a posteriori}. This would not be an ensemble approach. In contrast, we merge together all the time series in the specific time windows and compute directly the total transfer entropy.

Note that the need for a sufficient statistics is also strongly affected by the so-called curse of dimensionality. Indeed, in general the unbiased estimation of the transfer entropy requires the $k$-th and $l$-th reconstruction of Markov processes as explained above, but the number of data points needed for the correct sampling of PDFs scales non-linearly with the dimension, i.e. with both $k$ and $l$. 

Another crucial point is that the transfer entropy in eq. \eqref{transfer_entropy_1}, when the IF is absent, is equal to zero only theoretically. When the sample size is finite and the transfer entropy is empirically measured, a bias is always present, regardless the estimator we use. In this framework, a key question is whether or not the values found for the transfer entropy are statistically significant, especially if we do not know \textit{a priori} the underlying PDFs. To perform such a test, we need forming the null hypothesis $H_0$ that the IF is zero and the relevant distribution of the transfer entropy would be if $H_0$ was true. Practically it means that we need surrogate time series $\mathbf{Y}_{t-\tau}^{(l)}$ such that $p(X_t\vert \mathbf{X}_{t-1}^{(k)};\mathbf{Y}_{t-\tau}^{(l)})=p(X_t\vert \mathbf{X}_{t-1}^{(k)})$. In order to achieve this we create surrogate trials by only shuffling the source time series $\mathbf{Y}_{t-\tau}^{(k)}$ and leaving $X$ untouched. Indeed if $X$ was shuffled, any correlation would results to be destroyed and the Markov condition in eq. \eqref{generalMarkov}, i.e. our null hypothesis, may not be fulfilled anymore. Finally, a threshold confidence is fixed at 0.95 and the corresponding critical value of the transfer entropy $\hat{T}_{Y \rightarrow X}(\hat{t}, \tau)$ is computed in each window, so that if our measurements of the IF are greater than $\hat{T}_{Y \rightarrow X}(\hat{t}, \tau)$ we can argue statistical significance. For this preliminary study we use only two suggorates to fix the background of transfer entropy values.

\section{Data preparation}\label{sec:3}

In order to investigate the IF between external driving, auroral electrojet activity and ring current dynamics during magnetic storm events, we use the Super-MAG high-latitude index SML and low-latitude index SMR, which are a generalization of the traditional AL and SYM-H, respectively. They are, as usual, derived from deviations with respect to the average value of the horizontal (H) component of the geomagnetic field measured from a network of nearly-auroral/equatorial ground-based magnetometers \citep{gjerloev2009global}. The choice of Super-MAG indices is motivated by the fact that, since the auroral oval moves towards lower latitudes during severe magnetic storms, classical high-latitude geomagnetic indices have some limitations in estimating the correct value of auroral electrojet current intensity. In this framework the Super-MAG collaboration introduced the generalized AE-indices, i.e. SML, SMU and SME, computed using more than 300 different stations. Furthermore, the Super-MAG collaboration has 98 magnetometers in the range of latitudes currently used for constructing SYM-H and $D_{st}$ indices. So that, this sub-network is used to build up the SuperMAG equivalent of the ring current proxies, namely the SMR index \citep{newell2012supermag}.

In this framework, the typical fingerprint of a magnetic storm is monitored through the SMR index, which exhibits a sudden depression towards negative values. On the other hand, the polar substorm activity, i.e. the magnetic disturbance caused by the auroral electrojet current flowing in the auroral region, is investigated by means of the SML index, which is mainly representative of the geomagnetic tail dynamics \citep{gjerloev2004substorm, davis1966auroral, kamide2004physical}. 
Finally, we use the z-component of the IMF $B_z$ collected from OMNI database to infer the IF from the solar wind to internal magnetosphere-ionosphere system.

\begin{figure}[!h]
    \centering
    \includegraphics[width=12cm]{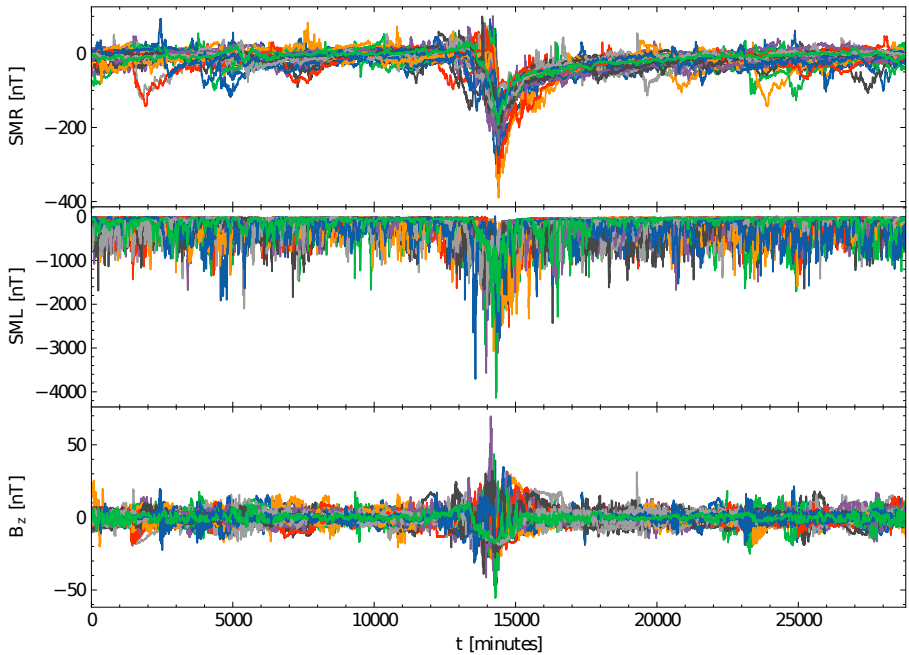}
    \caption{From top to bottom, SMR, SML and $B_z$. The time series are collected at 1 minute resolution.}
    \label{fig:1}
\end{figure}

The aforementioned ensemble of magnetic storms is now introduced. In detail, we started with a 23-year dataset (from 1995 to 2018) of $B_z$, SML and SMR from which we selected a set of magnetic storm periods for which $\text{SMR}\leq-150$ nT by considering a period of 10 days before and after the minimum of SMR during each storm event. In order to complete the ensemble we use the same periods of time for SML and $B_z$. As a last step, the double peaked storms have been removed from the ensemble by visual inspection since such complex events could introduce spurious effects when considered in our ensemble-based analysis.
The final dataset consists of $N_r=30$ independent storms that are reported in Figure \ref{fig:1}.

\section{Results}\label{sec:4}
\begin{figure}[t]
    \centering
    \includegraphics[width=10cm]{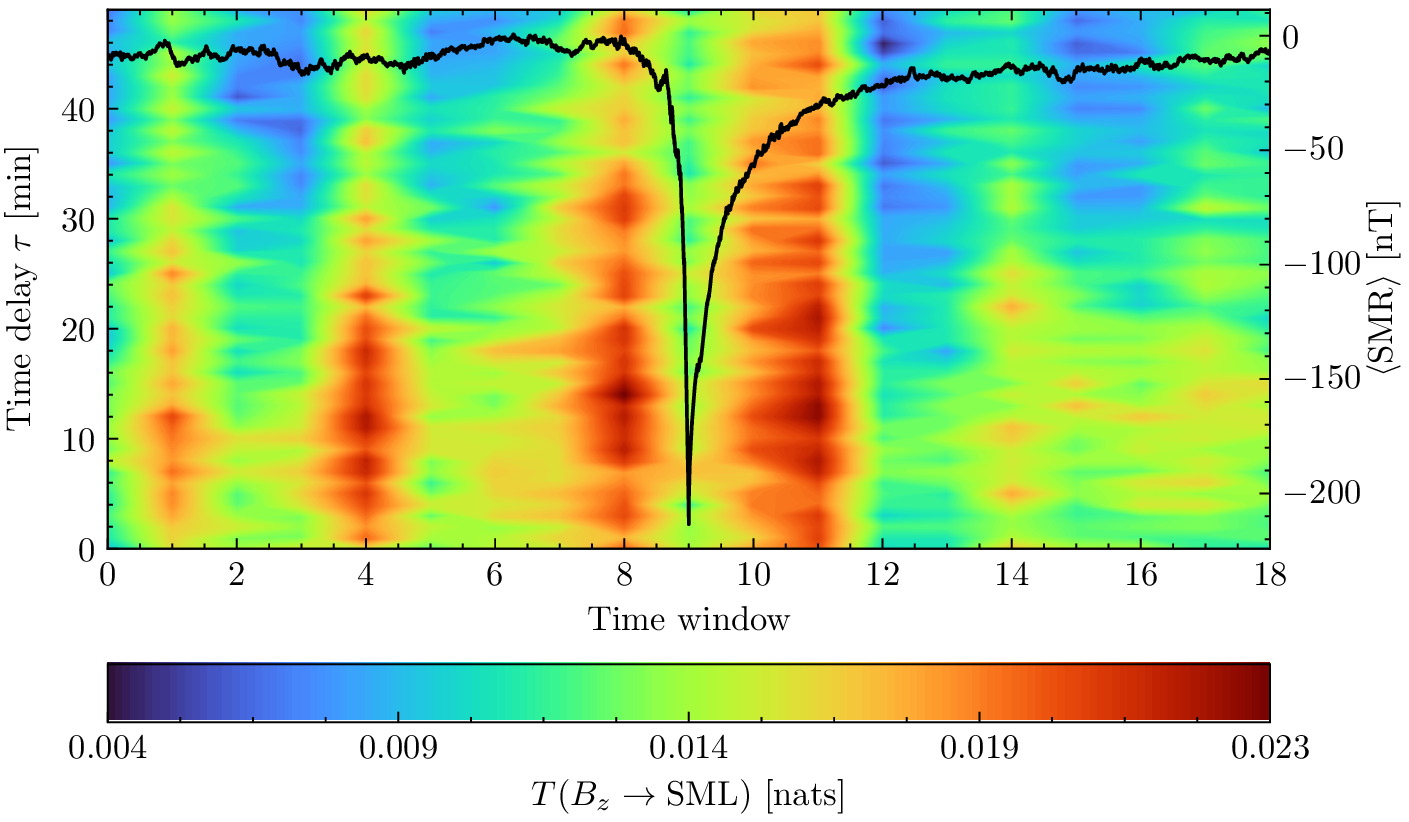}

    \includegraphics[width=10cm]{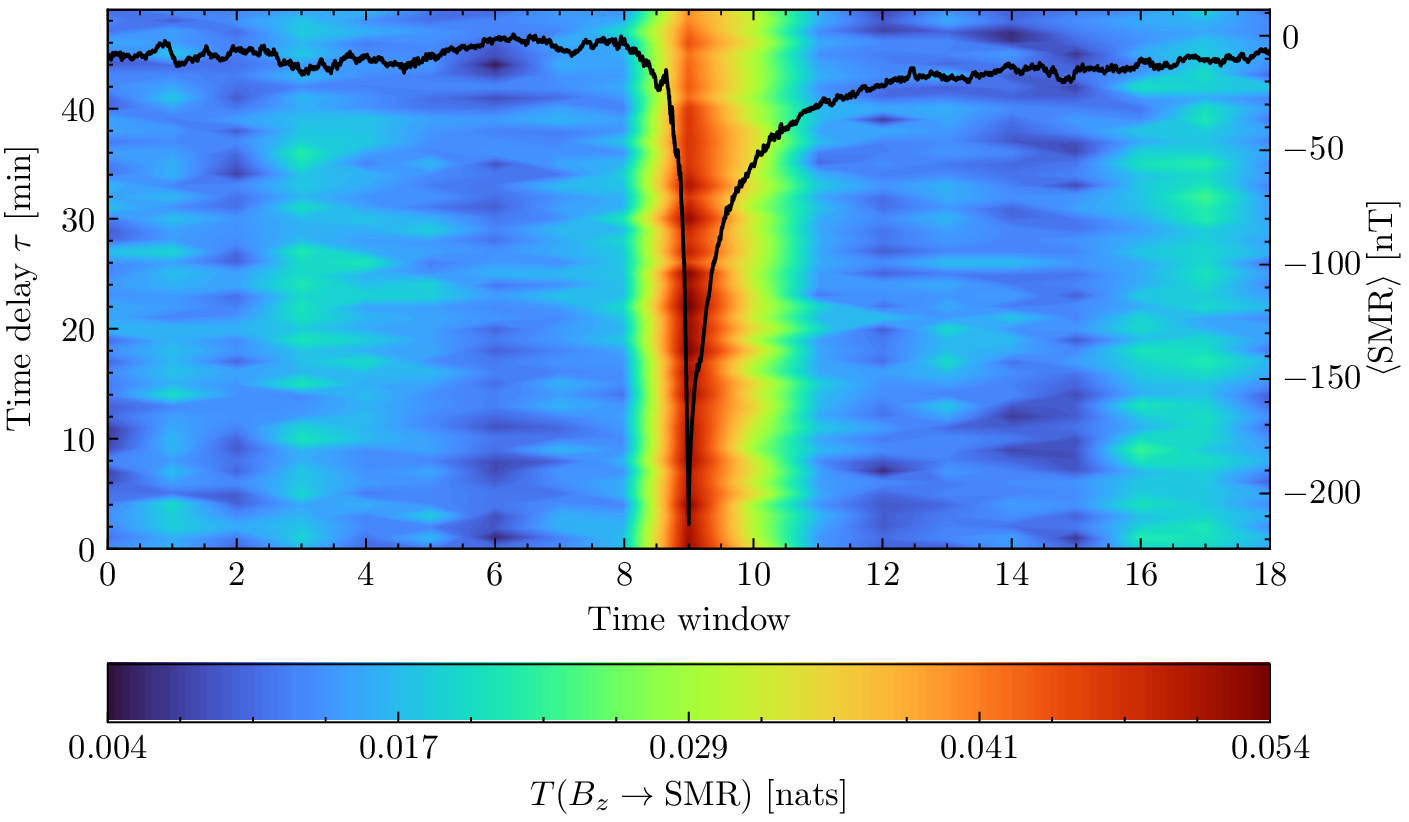}
    \caption{\textit{Top:} Contour plot of the transfer entropy from $B_z$ to SML with respect to the time window and to the time delay $\tau$. In order to compare the IF in terms of storm phases, the averaged track of SMR is depicted in black solid-line. \textit{Bottom:} Contour plot of the transfer entropy from $B_z$ to SMR with respect to the time window and to the time delay $\tau$. In order to compare the IF in terms of storm phases, the averaged track of SMR is depicted in black solid-line.}
    \label{fig:2}
\end{figure}
The main objective of this work is to provide a novel approach to inspect the IF within impulsive and strongly non-stationary processes, such as magnetic storms and magnetospheric substorms. As a first step we aim to evaluate the IF from the $z$-IMF component, which is representative of the driver, to both high-latitude and low-latitude geomagnetic activity by means of SML and SMR indices, respectively. The top panel of Figure \ref{fig:2} shows the contour plot of the ensemble-based TE $T(B_z\to\text{SML})$ as a function of the time window and the time delay $\tau$. We report the ensemble-averaged trend of the SMR index inside the figures, since this index is the one we used in the event synchronization and moreover it serves as a guide for the eye in identifying all the different phases of the magnetic storm. As is clear from the top panel of Figure \ref{fig:2}, there are different enhancements in the IF from $B_z$ to SML that are not related to the occurrence of the magnetic storm. Furthermore, the highest values of the TE are reached in proximity of the storm, i.e. during the pre-storm period and within the recovery phase, whereas a sudden decrease in the IF is observed during the storm main phase. Conversely, if we consider $T(B_z\to\text{SMR})$, reported in the bottom panel of Figure \ref{fig:2}, a significant enchancement of the IF is only present during the storm main phase. In this framework, these results emphasize the different role that the driver ($B_z$ in this case) plays in contributing to the dynamics of storms and substorms.
\begin{figure}[t]
    \centering
    \includegraphics[width=10cm]{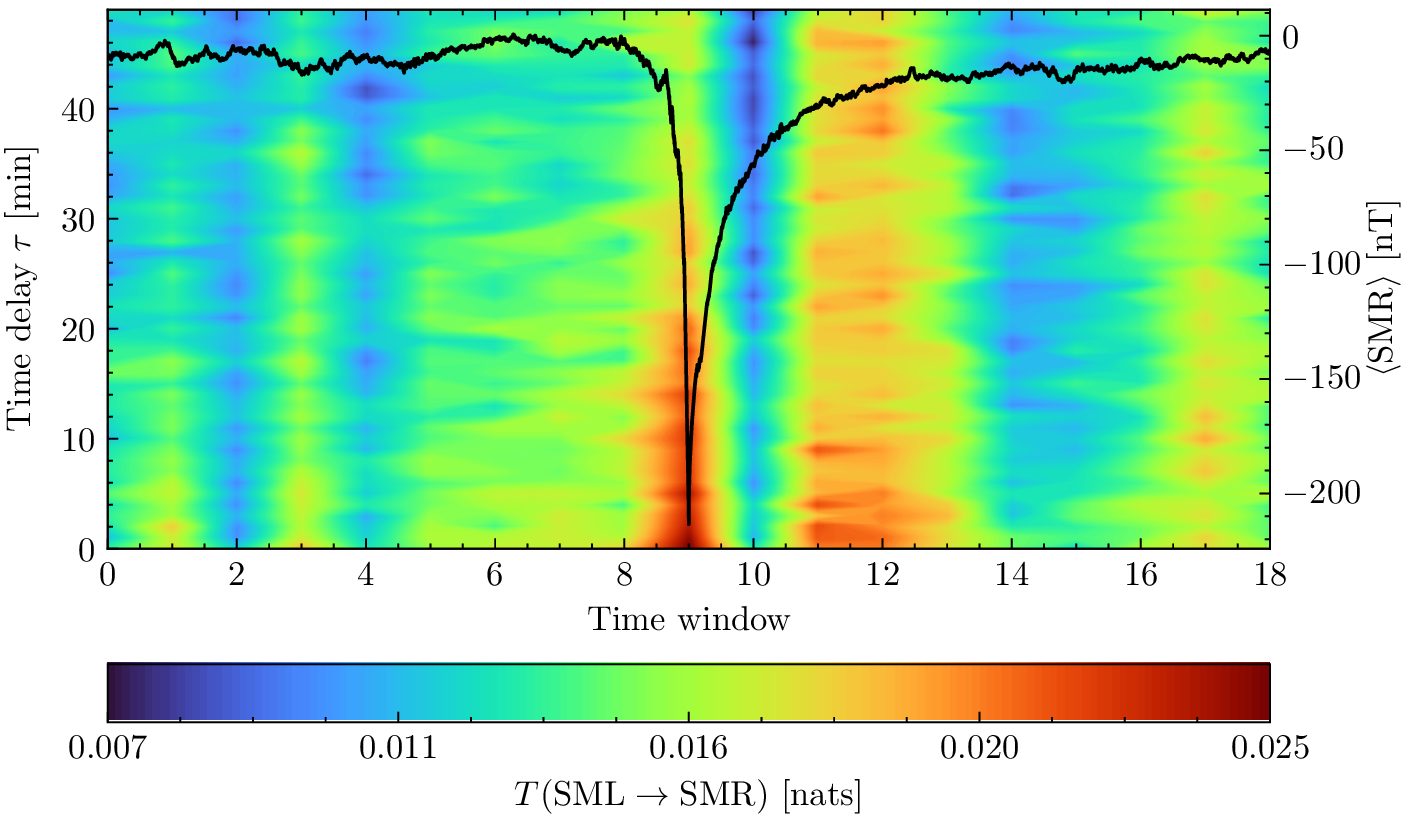}

    \includegraphics[width=10cm]{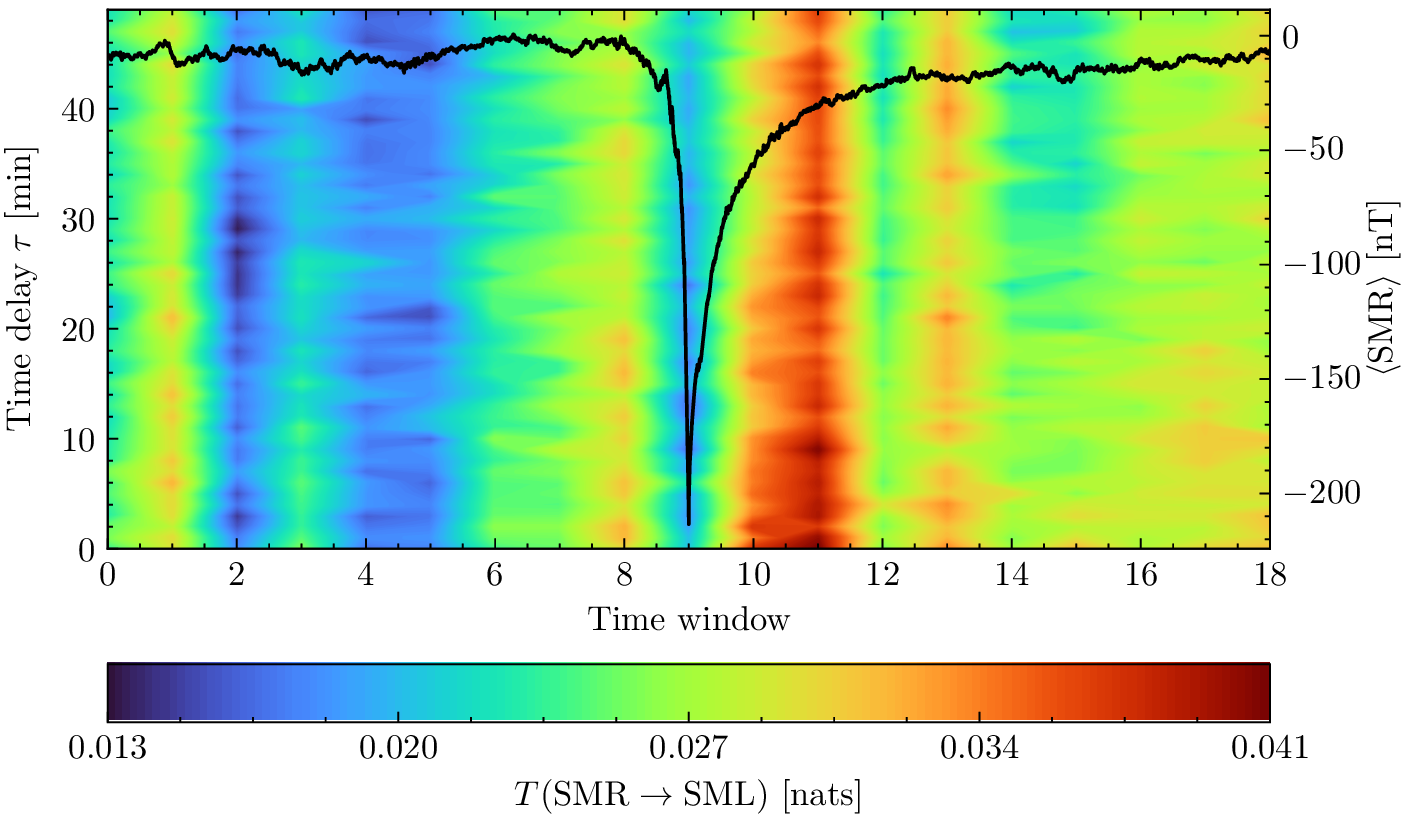}
    \caption{\textit{Top:} Contour plot of the transfer entropy from SML to SMR with respect to the time window and to the time delay $\tau$. In order to compare the IF in terms of storm phases, the averaged track of SMR is depicted in black solid-line. \textit{Bottom:} Contour plot of the transfer entropy from SMR to SML with respect to the time window and to the time delay $\tau$. In order to compare the IF in terms of storm phases, the averaged track of SMR is depicted in black solid-line.}
    \label{fig:3}
\end{figure}

In terms of internal dynamics, the characterization of the IF within the magnetosphere-ionosphere system is in general much more complex, since such flow is not unidirectional and feedback processes may be present as well. In this case it is crucial to elucidate the dynamics of the IF. In previous works, the investigation of the IF within the magnetosphere-ionosphere system has been carried by using a one-year dataset and the TE has been estimated over the whole signals neglecting possible time variations of the IF, e.g. in terms of intensity or direction. 
The ensemble TE from SML to SMR and vice-versa are reported, respectively, in the top and bottom panels of Figure \ref{fig:3}. By looking at $T(\text{SML}\to\text{SMR})$ it is clear how the maximum transfer of information from SML towards SMR is strongly localized around the minimum of $\langle \text{SMR}\rangle$, i.e. during the storm main phase. Moreover, the time lag $\tau$ at which the maximum is located is $\sim0$. By looking at $T(\text{SMR}\to\text{SML})$, which is representative of the IF from the ring current to the westward auroral electrojet current system, we observe a quite different scenario. The first enhancement of the TE approaching the storm is located just before the depression of SMR, whereas the maximum TE values are reached in the recovery phase. Contrary to what is observed for $\text{SML} \rightarrow \text{SMR}$, a sudden decrease of the IF between SMR and SML is observed at the onset of the mean main phase.


\section{Discussion and conclusion}\label{sec:5}

In this study we provided a first attempt to characterize the dynamics of the IF within the magnetosphere-ionosphere system using a database of magnetic storms instead of considering a long time series of geomagnetic indices. This allows us to avoid mixing the statistics of quiet and disturbed periods, as well as, thanks to our moving-window approach to follow the transition from quiet and disturbed conditions. However, one of the main limitation in considering the IF as an intrinsically non-stationary measure during transient periods, is the need for a sufficient statistics which clearly cannot be guaranteed. In order to overcome this problem, we introduced the analysis of transfer entropy over an ensemble of independent realizations of magnetic storms in a way similar to the method proposed by Gómez-Herrero et al. \citep{gomez2015assessing}. We emphasize again that this approach is somewhat different from computing the transfer entropy between individual trials and then averaging the single results \textit{a posteriori}.

We presented our approach by analyzing an ensemble of 30 independent magnetic storms. Firstly we studied the dynamics of the IF from solar wind to the magnetosphere-ionosphere system and we found a delayed information transfer according to previous findings \citep{de2011information, alberti2017timescale, stumpo2020measuring, runge2018common, manshour2021causality}. However, whereas the IF from $B_z$ to SMR, i.e., from the solar wind to the low-latitude magnetosphere (ring current), is enhanced only during the onset of the main phase of a magnetic storm, the IF from $B_z$ to SML, i.e., from the solar wind to the polar ionosphere, enhances not only during storm-times. This fact may be explained by the fact that auroral disturbance, i.e. magnetospheric substorms, can also occur outside a magnetic storm \citep{kamide1992substorm}. Indeed, whereas the development of a main-phase requires a southward oriented $B_z$ for a sufficient long time, the injection of solar wind particles into the polar ionosphere, i.e., the onset of geomagnetic tail reconnection and the successive impulsive energy dissipation through magnetospheric substorms, occur whenever $B_z$ is southward-oriented. 

The study of the internal dynamics, i.e., the magnetosphere-ionosphere coupling, is in general much more complex to interpret because a large number of current systems are involved. In this framework we found an important IF from SML to SMR at the beginning of the depression of SMR, which can be interpreted as the contribution of the outflow from the ionosphere. The current systems which act as mediators between the magnetosphere and the ionosphere in this case are the Field-Aligned-Currents (FACs). When the minimum of the disturbance is reached, the IF drops abruptly and is enhanced again during the recovery phase. The coupling induced by the FACs is not unidirectional, indeed they form a closed system with a reverse IF from SMR to SML, especially during the recovery phase. A possible explanation is that the excess of energetic particles are re-injected into the ionosphere from the ring current, where dissipation occurs via secondary substorms. On the other hand, this effect may be due to the non-Markovian nature of SML index at time-scales larger than 60 minutes as recently demonstrated by Benella et al. \citep{benella2022markov}. Without an appropriate reconstruction of the $l$-th order Markov process (see Section \ref{sec:2}), limited essentially by the need for a sufficient statistics, the effects of non-Markovianity may not be negligible so that the IF may be overestimated due to SML itself in this case.
\begin{figure}[t]
    \centering
    \includegraphics[width=10cm]{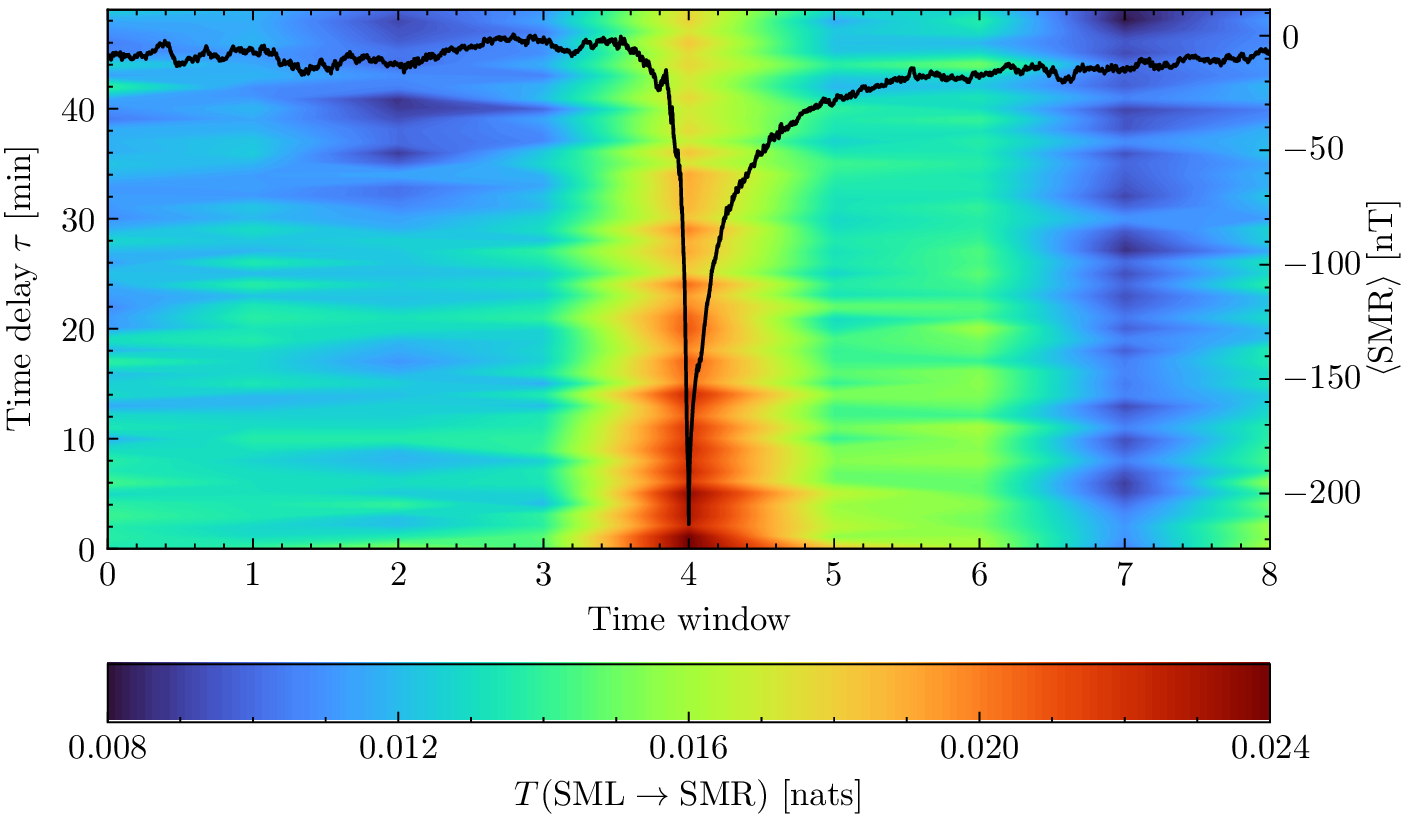}

    \includegraphics[width=10cm]{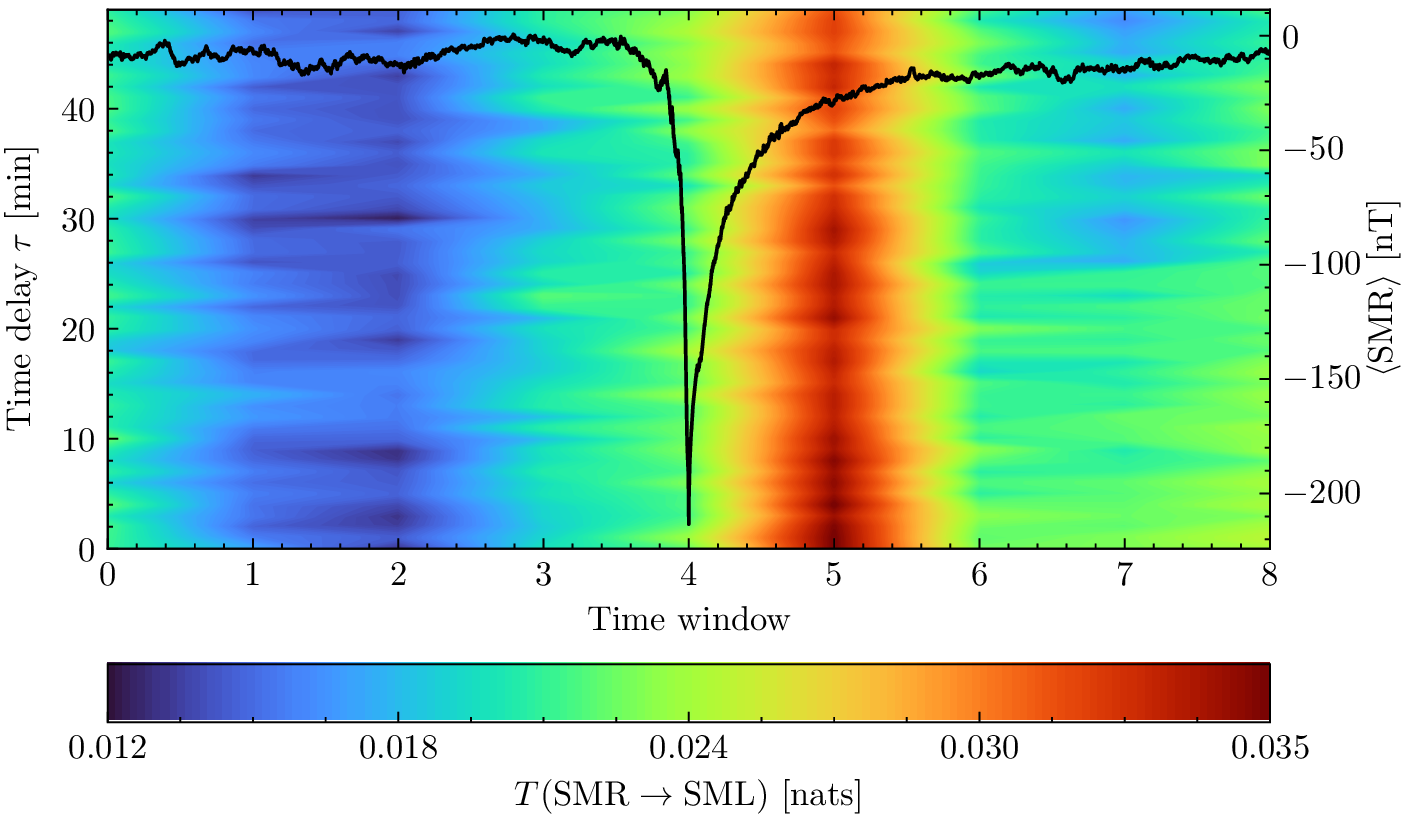}
    \caption{\textit{Top:} Contour plot of the transfer entropy from SML to SMR computed by using a window width of 1 day. \textit{Bottom:} Contour plot of the transfer entropy from SMR to SML compute by using a window width of 1 day.}
    \label{fig:4}
\end{figure}

The picture that during the early stages of a magnetic storm there exists a concurrent effect between a direct driving due to the solar wind activity and high-latitude processes is in agreement with the energy balance equation of $D_{st}$ index written in the form of eq. \eqref{energy_balance},
which enabled very good predictions of $D_{st}$ \citep{kamide1971analysis, gonzalez1994geomagnetic, kamide1998current}. In particular, eq. \eqref{energy_balance} represents the interplay between the direct external driver (i.e., efficiency of magnetic reconnection and efficiency of energy supply) and those internal processes triggered by the energy input given by the external driver. From a statistical point of view, our findings are also in great agreement with a very recent study by Alberti et al. \citep{alberti2022concurrent}. By using a novel approach based on dynamical systems theory, they computed the dimension of the reconstructed phase space by firstly considering AL and SYM-H alone and then by considering the joint process (AL, SYM-H). Interestingly, from this analysis figured out an independent contribution of AL in the dynamics of SYM-H during the development of the main-phase, in agreement with the IF between SML and SMR.

At this stage, it is also important to mention that the window width used for computing the transfer entropy in eq. \eqref{transfer_entropy_2}, naturally influences the behaviour of the IF. This is not surprising since the window width defines the time-scales in which the IF is measured. For example, if we compute the transfer entropy using a window width of 2 days, we found the results shown in Figure \ref{fig:4} for the IF from SML to SMR (top panel) and vice-versa (bottom panel). In this case we can see only the contribution of SML to the outflow localized just during the development of the main phase. In the reverse direction, i.e. from SMR to SML, we found a feedback process localized during the start of the recovery phase. Therefore, this behaviour highlights again the dependence of the IF on the time-scales in which it is measured. The time-scale dependence of the coupling for the case external-internal processes has been highlighted by Alberti et al. \citep{alberti2017timescale} by using the delayed mutual information on the filtered signals.

The works by Runge et al. \citep{runge2018common} and Manshour et al. \citep{manshour2021causality}, in contrast to previous findings by De Michelis et al. \citep{de2011information} and Stumpo et al. \citep{stumpo2020measuring}, found that the IF from the high-latitude to low-latitude (and vice-versa) is completely explained by the IMF, which might be the common driver. However, these results must be carefully interpreted since they provide an average view of the SMI system. This is related to the use of long time series to infer the IF despite the fact that magnetic storms and substorms do represent transient dynamics of the magnetosphere-ionosphere system. 
Furthermore, the conditional transfer entropy used by Manshour et al. \citep{manshour2021causality} is implicitly averaged for both positive and negative values of $B_z$, i.e., without providing any discrimination between open and closed conditions of the magnetosphere. Furthermore, this approach does not take into account preconditioning features of the magnetosphere-ionosphere system. From a phenomenological point of view and again with the help of eq. \eqref{energy_balance}, it means that periods when the coupling function of the magnetosphere-ionosphere systems is virtually set to zero (closed magnetosphere) are averaged together with those periods in which the coupling function is considerably different from zero (open magnetosphere). The relative importance of these two contributions is dependent on the total time the conditions explained above are satisfied, so that non-storm time coupling dominates the time average of the IF. This argument may explain the absence of the IF found by Manshour et al. \citep{manshour2021causality} and Runge et al. \citep{runge2018common} and, of course, the reason why we performed the analysis without removing the past-history of $B_z$. A more comprehensive analysis including the difference of southward and northward periods will be presented in a forthcoming paper.

In conclusion, our method provides a framework to study the time-variations of the IF at fixed time-scale. It is particularly suitable for the study of the relation between magnetic storms and substorms and, of course, of the magnetosphere-ionosphere coupling. Nevertheless, in this preliminary study some technical problems such as the reconstruction of l-th and k-th order Markov process as well as the accurate computation of the statistical threshold, have been considered only qualitatively. To assess  must be followed by a more comprehensive analysis. From physics side, it is interesting to discriminate the IF during northward (closed magnetosphere) and southward (open magnetosphere) IMF periods separately. This may reveal some interesting features of the injection/energization processes as well as the importance of the solar wind dynamic pressure, solar wind velocity and convection electric field. This is only the starting point. For accounting all this ideas, further work is needed. We remark also that in this preliminary study we used SuperMAG data. It will be interesting to perform the same analysis on SYM-H and AL indices. Again, this is matter of a more comprehensive work which will be given in future.

\backmatter

\bmhead{Acknowledgments}
The results presented in this paper rely on data collected at SuperMAG. We gratefully acknowledge the SuperMAG collaborators (\url{https://supermag.jhuapl.edu}).
The OMNI data were obtained from the Space Physics Data Facility (SPDF) Coordinated Data Analysis Web (CDAWeb) interface at \url{https://cdaweb.gsfc.nasa.gov/ index.html/}.
M.S. acknowledges the PhD course in Astronomy, Astrophysics and Space Science of the University of Rome “Sapienza”, University of Rome “Tor Vergata” and Italian National Institute for Astrophysics (INAF), Italy. G.C. and S.B. acknowledge the financial support by Italian MIUR-PRIN grant 2017APKP7T on Circumterrestrial Environment: Impact of Sun-Earth Interaction.

\bibliography{InfoFlow.bib}
\end{document}